\documentclass[manuscript]{aastex631}
\usepackage{amsmath}

\usepackage{CJKutf8}

\shorttitle{AASTeX v6.3.1 Sample article}
\shortauthors{Wang et al.}

\graphicspath{{./}{figures/}}
\begin{document} 
\begin{CJK}{UTF8}{gbsn}
\title{Search for Classical Cepheids in Galactic Open Clusters and Calibration of the Period--Wesenheit--Metallicity Relation in the Gaia Bands}

\author{Huajian Wang}
\affiliation{Purple Mountain Observatory, Chinese Academy of Sciences, Nanjing 210008, Peopleʼs Republic of China; xuye@pmo.ac.cn}
\affiliation{University of Science and Technology of China, 96 Jinzhai Road, Hefei 230026, Peopleʼs Republic of China}
\author{Ye Xu}
\affiliation{Purple Mountain Observatory, Chinese Academy of Sciences, Nanjing 210008, Peopleʼs Republic of China; xuye@pmo.ac.cn}
\affiliation{University of Science and Technology of China, 96 Jinzhai Road, Hefei 230026, Peopleʼs Republic of China}

\author{Zehao Lin}
\affiliation{Purple Mountain Observatory, Chinese Academy of Sciences, Nanjing 210008, Peopleʼs Republic of China; xuye@pmo.ac.cn}
\affiliation{University of Science and Technology of China, 96 Jinzhai Road, Hefei 230026, Peopleʼs Republic of China}

\author{Chaojie Hao}
\affiliation{Purple Mountain Observatory, Chinese Academy of Sciences, Nanjing 210008, Peopleʼs Republic of China; xuye@pmo.ac.cn}
\affiliation{University of Science and Technology of China, 96 Jinzhai Road, Hefei 230026, Peopleʼs Republic of China}

\author{Dejian Liu}
\affiliation{Purple Mountain Observatory, Chinese Academy of Sciences, Nanjing 210008, Peopleʼs Republic of China; xuye@pmo.ac.cn}
\affiliation{University of Science and Technology of China, 96 Jinzhai Road, Hefei 230026, Peopleʼs Republic of China}

\author{Yingjie Li}
\affiliation{Purple Mountain Observatory, Chinese Academy of Sciences, Nanjing 210008, Peopleʼs Republic of China; xuye@pmo.ac.cn}



\begin{abstract}
It is beneficial to calibrate the period--Wesenheit--metallicity relation (PWZR) of Delta Cephei stars (DCEPs), i.e., classical Cepheids, using accurate parallaxes of associated open clusters (OCs) from Gaia data release 3 (DR3). To this aim, we obtain a total of 43 OC-DCEPs (including 33 fundamental mode, 9 first overtone mode, and 1 multimode DCEPs.) and calibrate the PWZR as $W_G=(-3.356 \,\pm\, 0.033) \,(\log{P-1})+(-5.947 \,\pm\, 0.025)+(-0.285 \,\pm\, 0.064)[\textrm{Fe/H}]$. The concurrently obtained residual parallax oﬀset in OCs, $zp = -4\pm5\,\mu\textrm{as}$, demonstrate the adequacy of the parallax corrections within the magnitude range of OC member stars. By comparing the field DCEPs' DR3 parallaxes with their photometric parallaxes derived by our PWZR, we estimated the residual parallax oﬀset in field DCEPs as $zp = -15\pm3\,\mu\textrm{as}$. Using our PWZR, we estimate the distance modulus of the Large Magellanic Cloud to be $18.482 \,\pm\, 0.040$ mag, which aligns well with the most accurate published value obtained through geometric methods.
\end{abstract}



\section{Introduction} \label{sec:intro}
Delta Cephei stars (DCEPs), i.e., classical Cepheids, constitute Population I of Cepheids and are located in the instability strip above the main sequence in color--magnitude diagrams \citep[CMDs;][]{turner06}. A DCEP is a youthful, periodic pulsating yellow giant or supergiant that has been around for tens to hundreds of millions of years, with a pulsation period of approximately 1 to 100 days. The period--luminosity relation, also known as Leavitt's law \citep{leavitt12}, is a well-known characteristic of DCEPs. Accurate distances derived from the period--luminosity relationship of DECPs are widely used. For example, DCEPs have been used to study the structure of the Milky Way \citep[e.g.,][]{chen19,skowron19}, measure the distances to other galaxies \citep[e.g.,][]{freedman01,sandage06}, and serve as a critical step in measuring the Hubble constant $H_0$ \citep[e.g.,][] {freedman11,riess21}.

Various reddening effects are produced by the different lines of sight and distances to DCEPs in the Milky Way. The period--Wesenheit relation (PWR) overcomes the limitations of reddening by transforming multiband magnitudes into Wesenheit magnitudes \citep{madore82,Majaess08}. Traditionally, the PWR of DCEPs is calibrated using the parallaxes of individual stars 
\citep[e.g.,][]{ripepi19,poggio21,ripepi22}. However, this method suffers from uncertain residual parallax offset in the Gaia parallaxes of the individual stars \citep{lindegren21}. Besides, it is predicted that a variation in metal abundance affects the shape and width of the DCEP instability strip \citep[e.g.,][]{caputo2000}, which consequently affects the coefficients of the PWR \citep[][and references therein]{mar05,mar10,desomma22}. Limitations in parallax accuracy led to a stagnation in studies of the period--Wesenheit--metallicity relation (PWZR) until the advent of the Gaia mission \citep{Collaboration16}, which has provided accurate parallaxes for a total of 1.8 billion objects to date, resulting in a large number of works on the PWZR to spring up \citep[e.g.,][]{greo18,ripepi19,ripepi20,ripepi21,riess21,breuval22,ripepi22,reyes23,trentin24}.

An alternative method to calibrating the PWR or PWZR is to use the parallaxes of open clusters (OCs) harboring DECPs. This newly developed method takes advantage of the stars in the OCs all having a similar distance, extinction, age, and metallicity, as well as the fact that the age distribution of OCs \citep[ranging from several million years to several billion years;][]{khar2013} partially overlaps with the age range of DECPs. After parallax corrections \citep[][hereafter L21]{lindegren21}, this method has been proven to eliminate residual parallax offset \citep{riess22,reyes23}, resulting in more accurate PWRs and PWZRs \citep[e.g.,][]{breuval20,zhou21,lin22,riess22,reyes23}. For example, \cite{riess22} used 17 OC-DCEPs with Hubble Space Telescope (HST) photometry to calibrate the PWZR in the HST photometric system and determine a precise Hubble constant of $H_0 = 73.01 \pm 0.99$ km {s}$^{-1}$ Mpc$^{-1}$.

The first OC-DCEP was discovered by \cite{doig1925}, and searches for OC-DECPs have been active in the past decade \citep{anderson13,chen15,clark15,chen17,lohr18,alonso20,breuval20,negueruela20,med21,zhou21,hao22b,lin22,reyes23}. Recently, there have been new searches for OCs \citep[e.g.,][]{hunt23} based on Gaia data release 3 \citep[DR3;][]{Collaboration23}. In this current study, by cross-matching Gaia sources with the 3655 Galactic DCEPs compiled by \cite{pie21}, we assemble a larger sample of OC-DECPs, which allows us to derive more accurate PWZRs.

The structure of this paper is arranged as follows. In Section \ref{sec:Data}, we introduce our extended OC-DECP sample. In Section \ref{sec:analysis}, we describe the calibration results for the PWZR derived with our samples. In Section \ref{sec:discussion}, we test the reliability of our PWZR on Galactic field DCEPs and Large Magellanic Cloud (LMC) field DCEPs. Finally, in Section \ref{sec:conclusions}, we summarize this work.

\section{Data} \label{sec:Data}

\subsection{Open Clusters} \label{sec:OCs}
There are a total of 7167 clusters in the \cite{hunt24} catalog, which covers almost all previously published OCs. Among these 7167 clusters, we only utilize 3530 high-quality OCs, which all identify clear isochrones by network training methods and filter out moving groups by Jacobi radius. The OCs'coordinates, proper motions, and parallaxes were extracted from \cite{hunt24}. It should be noted that OC parallaxes from \cite{hunt24} were derived through the maximum likelihood distances, where L21 corrections had been considered. We obtained the OC parallax error $\sigma_{\varpi_{\textrm{OC}}}$ by the quadrature sum of the statistical uncertainty and the angular covariance\footnote{Due to the large number of member stars in an OC, the statistical uncertainty of the OC's 
parallax will benefit from the $\sqrt{N}$ improvement. However, as
the angular covariance of the Gaia parallaxes is much larger \citep{lindegren21,ape21,vas21,zin21}, we took into account the angular covariance.} defined by \cite{ape21}.

\subsection{Classical Cepheids} \label{sec:DCEPs}

\cite{pie21} compiled a sample of 3,655 DCEPs in the Milky Way. They also supplied the DR3 source\_id of each DCEP by applying a matching radius of 0.''5. We used their DR3 source\_id to match the gaiadr3.gaia\_source and extracted the required parameters (e.g., $\alpha$, $\delta$, $\varpi$, $\mu_{{\alpha}^\ast}$, and $\mu_\delta$) for each DCEP. To obtain more reliable photometry of DCEPs, we extracted the intensity-averaged magnitudes ($m_{G_{\textrm{BP}}}$, $m_{G_{\textrm{RP}}}$, and $m_{G}$) from gaiadr3.vari\_cepheid\footnote{\url{https://gea.esac.esa.int/archive/}} for 3,046 DCEPs.

\subsection{Cross match} \label{sec:match}
OC-DCEPs are identified if the following criteria are met: (1) The projected distance between DCEPs and OCs should be less than 25 pc, assuming that the parallaxes of DCEPs are equal to those of OCs. (2) The $\mu_{{\alpha}^\ast}$, $\mu_\delta$, and $\varpi$ of DCEPs should be within 3$\sigma$ ($\sigma$ is the standard deviation of OC) of those of OCs. Additionally, an expanded sample is taken into account, in which a few dimensions are slightly higher than 3$\sigma$ but less than 3.5$\sigma$. (3) DCEPs should be located on the instability strip of their host OC’s CMD \citep{turner06}. After filtering using the above criteria, we obtained 43 OC-DCEPs, whose astrometry and photometry are given in Table \ref{tab:2}. Representative examples of OC-DCEPs and rejected associations are shown in panels (a) and (b) of Figure \ref{fig:6}, respectively. It should be noted that among the 43 OC-DCEPs we obtained, there is an association of U~Sgr with the OC IC~4725, but U~Sgr is not in gaiadr3.vari\_cepheid and is hence not used for PWZR calibration.

\section{Analysis}\label{sec:analysis}
Our 43 OC-DCEPs are composed of 33 DCEPs pulsating in the fundamental mode (F-mode), nine pulsating in the first overtone (1O-mode), and one multimode (F1O-mode) pulsator. To establish the PWZR of the DCEPs including the 1O-mode DCEPs, we used the equation $P_{\textrm{F}} = P_{\textrm{1O}}/(0.716-0.027\log P_{\textrm{1O}})$ \citep{feast97} to obtain their period in the F-mode, where $P_{\textrm{F}}$ and $P_{\textrm{1O}}$ represent pulsations in the F-mode and 1O-mode, respectively. For that F1O-mode DCEP, we adopted its period in the F-mode.

To obtain the metal abundances of our OC-DCEPs, we matched our OC-DCEPs with \cite{trentin24}, who compiled 910 DCEPs with literature metal abundances from high-resolution spectroscopy or metal abundances from the Gaia Radial Velocity Spectrometer (see Section 2.2 in \citealp{trentin24}, for details). Finally, we obtained the metal abundances of 40 OC-DCEPs and compiled them in Table \ref{tab:2}.

To calibrate the PWZR in the Gaia bands, we refer to the method in \cite{riess22} and \cite{ripepi22}. The photometric parallax (in milliarcseconds) is defined as:

\begin{equation}
\varpi_{\textrm{phot}} = 10^{-0.2(w_G - W_G - 10)},
\end{equation}
where $w_G$ is the apparent Wesenheit magnitude and can be defined as $w_G = m_G - \lambda \times(m_{G_{\textrm{BP}}} - m_{G_{\textrm{RP}}})$. We adopted the empirical result $\lambda = 1.9$ \citep{ripepi19}. $W_G$ is the absolute Wesenheit magnitude, which can be defined as:

\begin{equation}
W_G=\alpha (\log{P} - 1) + \beta + \gamma[\textrm{Fe/H}].
\end{equation}

We used the \texttt{optimize.minimize} method from the \texttt{Python Scipy} library to minimize the following quantity:
\begin{equation}
\chi^2 = \sum{\frac{\left(\varpi_{\textrm{OC}}-\varpi_{\textrm{phot}}+zp\right)^2}{\sigma^2}} = \sum{\frac{\left(\varpi_{\textrm{OC}}-\varpi_{\textrm{phot}}+zp\right)^2}{{\sigma_{\varpi_{\textrm{OC}}}^2 + \sigma_{\varpi_{\textrm{phot}}}^2 }}},
\end{equation}
where $zp$ is the residual parallax offset in OCs. For $\sigma_{\varpi_{\textrm{phot}}}$, we refer to the definition given in \cite{ripepi22}: $\sigma_{\varpi_{\textrm{phot}}} = 0.46 \times \sqrt{\sigma_{w_G}^2 + \sigma_{W_G}^2 } \times \varpi_{\textrm{phot}}$ and $\sigma_{w_G}$ is calculated by error propagation, assuming a conservative error of 0.02 mag for the three Gaia bands ($G_{\textrm{BP}}$, $G_{\textrm{RP}}$, and $G$). We adopted a conservative dispersion of 0.1 mag for $\sigma_{W_G}$ \citep{desomma20}.

\begin{figure*}[htb]
\gridline{\fig{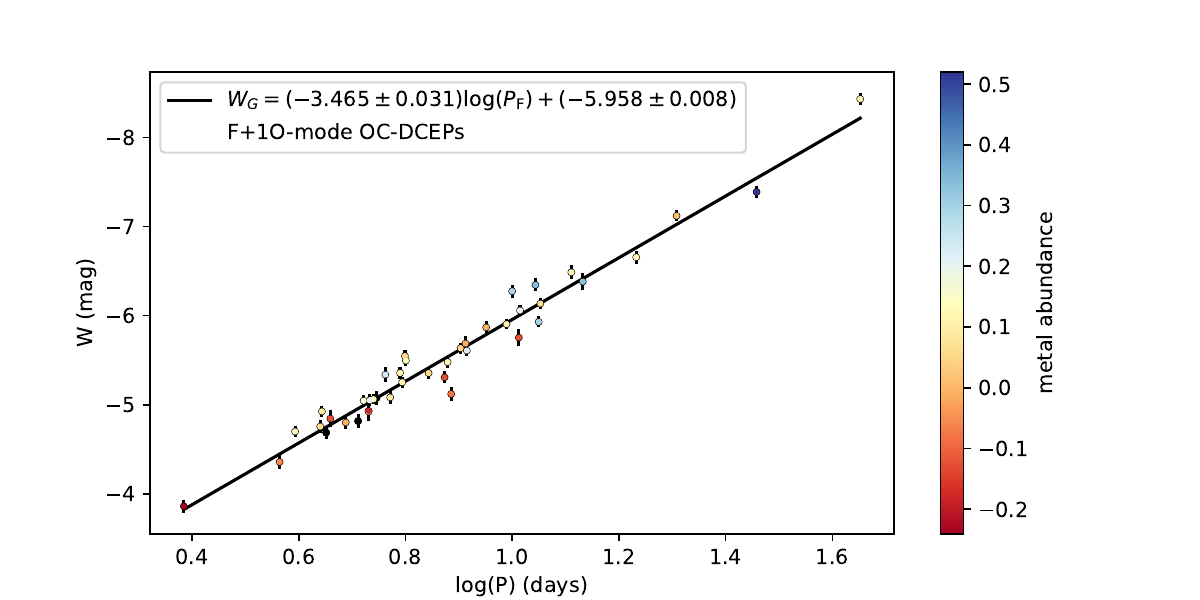}{0.5\textwidth}{} \hfill \fig{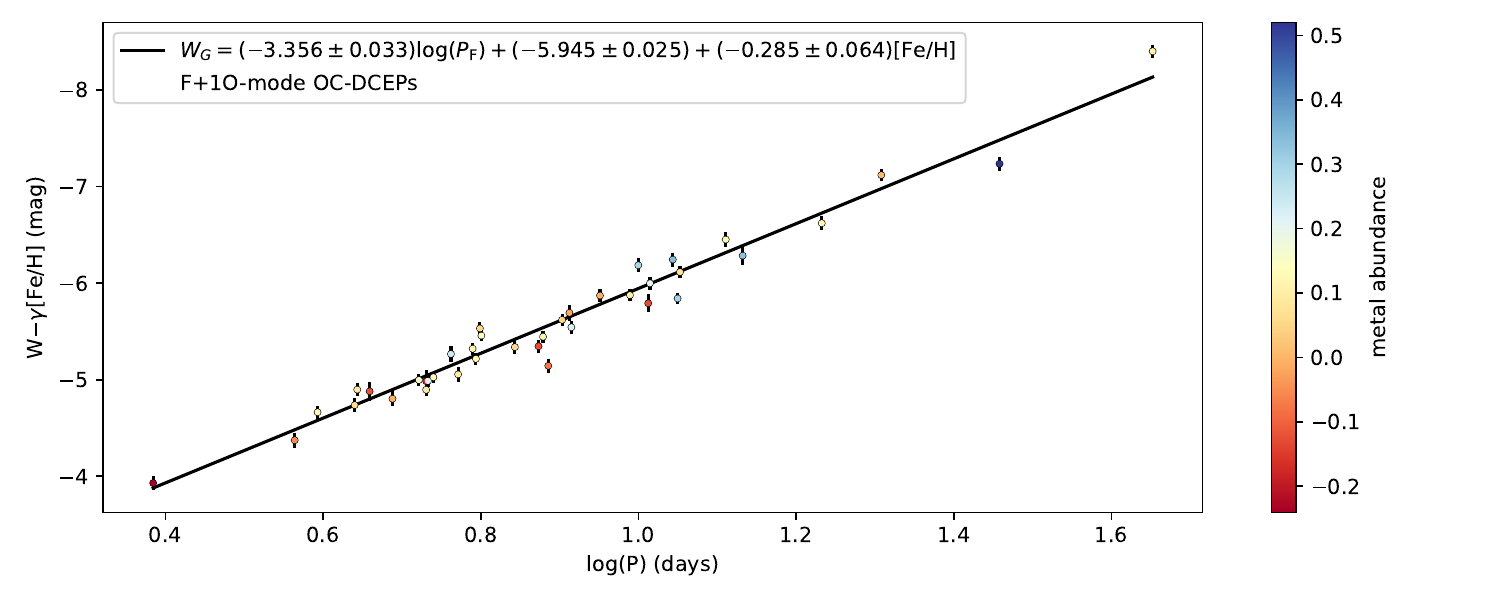}{0.57\textwidth}{}}
\gridline{\fig{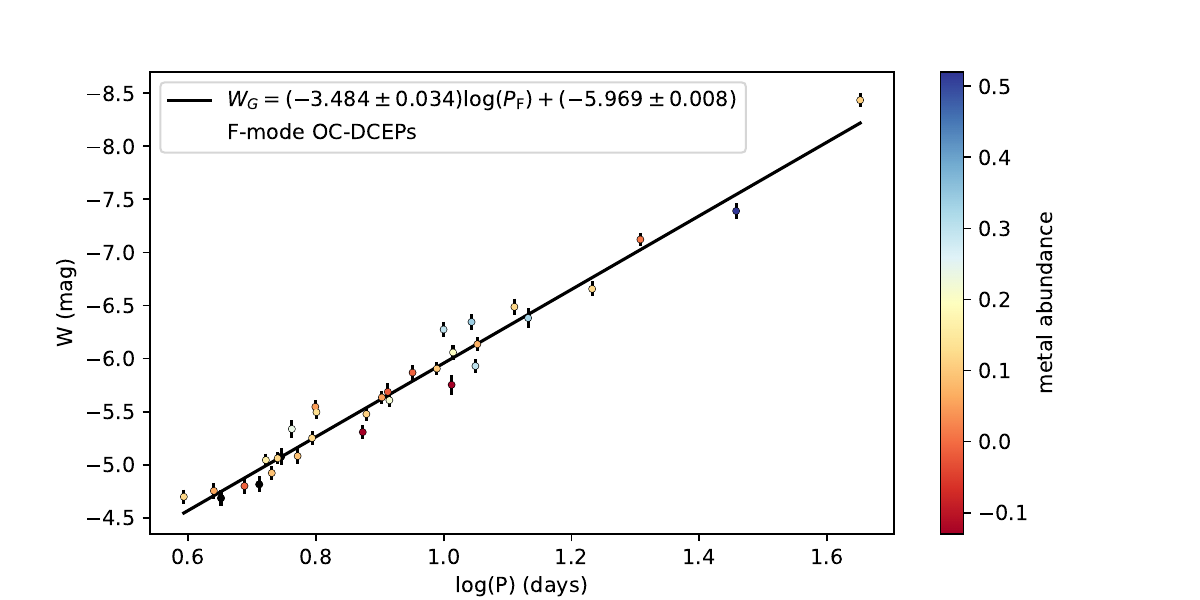}{0.5\textwidth}{} \hfill \fig{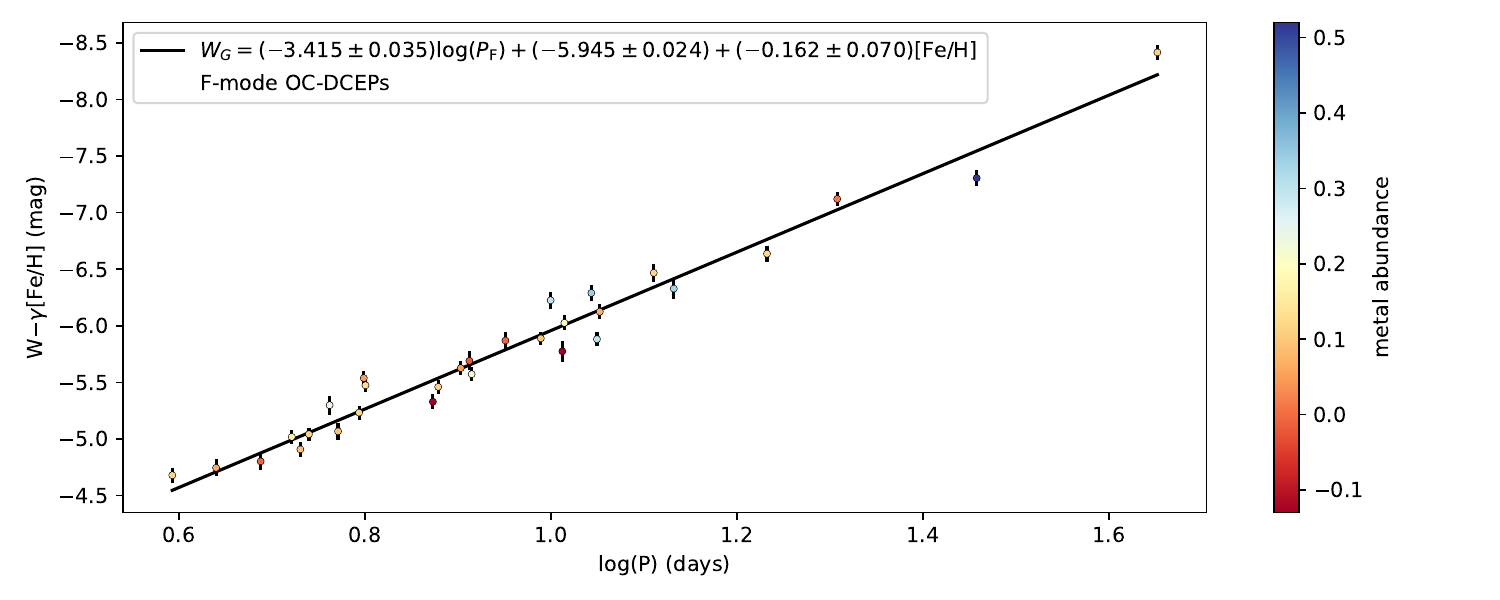}{0.57\textwidth}{}}
\caption{PWR and PWZR fitting results from our OC-DCEPs. The color represents the value of metal abundance. The black dots in the two subfigures on the left are OC-DCEPs without literature metal abundances.
\label{fig:1}}
\end{figure*}

To ensure the robustness of the fit, we performed 10,000 Monte Carlo simulations, where for each simulation we randomly varied $\varpi_{\textrm{OC}}$ and $\sigma_{w_G}$ within their errors to obtain the distribution of each coefficient. Each distribution's median and standard deviation are then taken as the best-fitting value of the coefficient and its error, respectively. The fitting results of our PWR and PWZR are shown in the left and right subfigures of Figure \ref{fig:1}, respectively. The marginalised posterior distributions of the free parameters in the fitting are shown in Figure \ref{fig:7}.

We present our PWZR and compare them with other works in Table \ref{tab:1}. The $zp$ of Case 3 and Case 4 in Table \ref{tab:1} are $-4\pm5\,\mu\textrm{as}$ and $1\pm5\,\mu\textrm{as}$, respectively, which proves the adequacy of L21 corrections within the magnitude range of OC member stars. As in most works, negative metallicity terms $\gamma$ are obtained. Specifically, our $\gamma$ of F-mode OC-DCEPs is $-0.162\,\pm\,0.070$, but the $\gamma$ of F+1O-mode OC-DCEPs is $-0.285\,\pm\,0.064$. This is consistent with the conclusions of \cite{desomma22}, who discovered the $\gamma$ of $\sim -0.1 \,\textrm{to}\sim-0.2$ for the F-mode and $\sim -0.1 \,\textrm{to}\sim-0.3$ for the F+1O-mode based on stellar pulsation models. We also found that the absolute value of the $\gamma$ obtained by the empirical relation of field DCEPs is larger than what we obtained. Our OC-DCEPs have a smaller range of metal abundances, which may explain the smaller absolute value of the $\gamma$ we obtained. In the future, obtaining more OC-DCEPs with a wider range of metal abundances will help us better constrain the $\gamma$.

\cite{reyes23} fixed the $\gamma$ to $-0.384 \,\pm\, 0.051$ \citep{breuval22} and then calibrated the PWZR in the Gaia bands using 26 F-mode OC-DCEPs and 225 field DCEPs as $W_G=(-3.242 \,\pm\, 0.044) \,\log(P - 1)+(-6.004 \,\pm\, 0.019)+(-0.384 \,\pm\, 0.051)[\textrm{Fe/H}]$. To compare with it, we adopted $\lambda=1.921$ and fixed the slope $\alpha$ to $-3.242 \,\pm\, 0.044$ and the $\gamma$ to $-0.384 \,\pm\, 0.051$. For our F+1O-mode and F-mode OC-DCEPs, the fitting results of intercept $\beta$ are $-6.008\,\pm\, 0.010$ mag and $-6.046\,\pm\, 0.010$ mag, respectively. Both fitting results are consistent with \cite{reyes23} within 3$\sigma$, and the error is smaller.

\begin{deluxetable*}{cccccccccc}[t]
\setlength{\tabcolsep}{3pt}
\tablenum{1}
\tablecaption{Comparison of the Gaia Bands PWZR Obtained in This Work with Other Works \label{tab:1}}
\tablewidth{0pt}
\tabletypesize{\footnotesize}
\tablehead{
\colhead{Case} & \colhead{$\alpha$}& \colhead{$\beta$}& \colhead{$\gamma$}&\colhead{$zp$}& \colhead{Mode}& \colhead{$\mu_{\textrm{LMC}}$}& \colhead{$\Delta\mu_{\textrm{LMC}}$}& \colhead{Sample}\\
\colhead{} &\colhead{} &\colhead{(mag)} &\colhead{} &\colhead{($\mu$as)} &\colhead{} &\colhead{(mag)} &\colhead{(mag)} &\colhead{}}

%\decimalcolnumbers
%\startlongtable
\startdata
\multicolumn{9}{c}{This work} \\ \hline
1 & $-3.465  \,\pm\,  0.031$ & $-5.958  \,\pm\,  0.008$ & -- & -- & F+1O & 18.557  $\,\pm\,$  0.014 & 0.080(5.7$\sigma$) & 42 OC-DCEPs \\ 
2 & $-3.484  \,\pm\,  0.034$ & $-5.969  \,\pm\,  0.008$ & -- & -- & F & 18.582  $\,\pm\,$  0.016 & 0.105(6.6$\sigma$) & 33 OC-DCEPs                  \\
3 & $-3.356  \,\pm\,  0.033$ & $-5.947  \,\pm\,  0.025$ & $-0.285  \,\pm\,  0.064$ & $-4\,\pm\,5\,\,\,\,\,$ & F+1O & 18.482  $\,\pm\,$  0.040 & 0.005(0.1$\sigma$) &  39 OC-DCEPs                  \\
4 & $-3.415  \,\pm\,  0.035$ & $-5.945  \,\pm\,  0.024$ & $-0.162  \,\pm\,  0.070$  & $1\,\pm\,5$ & F & 18.520  $\,\pm\,$  0.040 & 0.043(1.1$\sigma$) & 30 OC-DCEPs \\ \hline
\multicolumn{9}{c}{\cite{ripepi22}} \\ \hline
5 & $-3.176  \,\pm\,  0.044$ & $-5.988\,\pm\,  0.018$ & $-0.520  \,\pm\,  0.090$  &-- & F+1O & 18.513  $\,\pm\,$  0.046 & 0.036(0.8$\sigma$) & 435 DCEPs \\ 
6 & $-3.178  \,\pm\,  0.048$ & $-5.971\,\pm\,  0.017$ & $-0.661  \,\pm\,  0.077$  &-- & F & 18.439  $\,\pm\,$  0.041 & 0.038(0.9$\sigma$) & 372 DCEPs \\\hline
\multicolumn{9}{c}{\cite{breuval22}} \\ \hline
7 & $-3.338  \,\pm\,  0.012$ & $-5.959  \,\pm\,  0.025$ & $-0.384  \,\pm\,  0.051$ & -- & F & 18.474  $\,\pm\,$  0.033 & 0.003(0.1$\sigma$) & 2473 DCEPs \\\hline
\multicolumn{9}{c}{\cite{reyes23}} \\ \hline
8 & $-3.242  \,\pm\,  0.047$ & $-6.004  \,\pm\,  0.019$ & $-0.384  \,\pm\,  0.051$ & $-19\,\pm\,3\tablenotemark{$^{\ast}$}\,\,\,\,\,$ & F & 18.540  $\,\pm\,$  0.034 & 0.063(1.9$\sigma$) & 26 OC-DCEPs + 225 DCEPs \\ \hline
\multicolumn{9}{c}{\cite{bhar23}} \\ \hline
9 & $-3.21  \,\pm\,  0.07$ & $-5.94  \,\pm\,  0.03$ & $-0.33  \,\pm\,  0.16$ & -- & F & 18.530  $\,\pm\,$  0.078 & 0.053(0.7$\sigma$) &  64 DCEPs   \\\hline
\multicolumn{9}{c}{\cite{bhar24}} \\ \hline
10 & $-3.54  \,\pm\,  0.06$ & $-6.21  \,\pm\,  0.03$ & $-0.47  \,\pm\,  0.10$ & -- & F+1O & 18.577  $\,\pm\,$  0.059 & 0.100(1.7$\sigma$) &  60 DCEPs   \\\hline
\multicolumn{9}{c}{\cite{trentin24}} \\ \hline
11 & $-3.230  \,\pm\,  0.041$ & $-5.960\,\pm\,  0.018$ & $-0.573  \,\pm\,  0.066$  &-- & F+1O & 18.438  $\,\pm\,$  0.038 & 0.039(1.0$\sigma$) & 726 DCEPs \\ 
12 & $-3.245  \,\pm\,  0.055$ & $-5.917\,\pm\,  0.017$ & $-0.745  \,\pm\,  0.085$  &-- & F & 18.323  $\,\pm\,$  0.045 & 0.154(3.4$\sigma$) & 478 DCEPs \\
\enddata
\tablenotemark{--}{ means that this parameter does not join in the fitting as a free parameter.}

\tablenotemark{$^{\ast}$}{ Residual parallax offset in field DCEPs after L21 corrections.}
\tablecomments{$\alpha$, $\beta$, $\gamma$, and $zp$ are the slope, intercept, metallicity term, and residual parallax offset, respectively. $\mu_{\textrm{LMC}}$ is the distance modulus of the LMC derived from the PWZR. $\Delta\mu_{\textrm{LMC}}$ is the difference between $\mu_{\textrm{LMC}}$ measured by PWZR and $\mu_{\textrm{LMC}}$ measured by \cite{pie19}.}
\end{deluxetable*}

\section{Discussion}\label{sec:discussion}
\subsection{Reliability Testing of PWZR on Galactic Field DCEPs} \label{sec:field}
\begin{figure}[htb]
\gridline{\fig{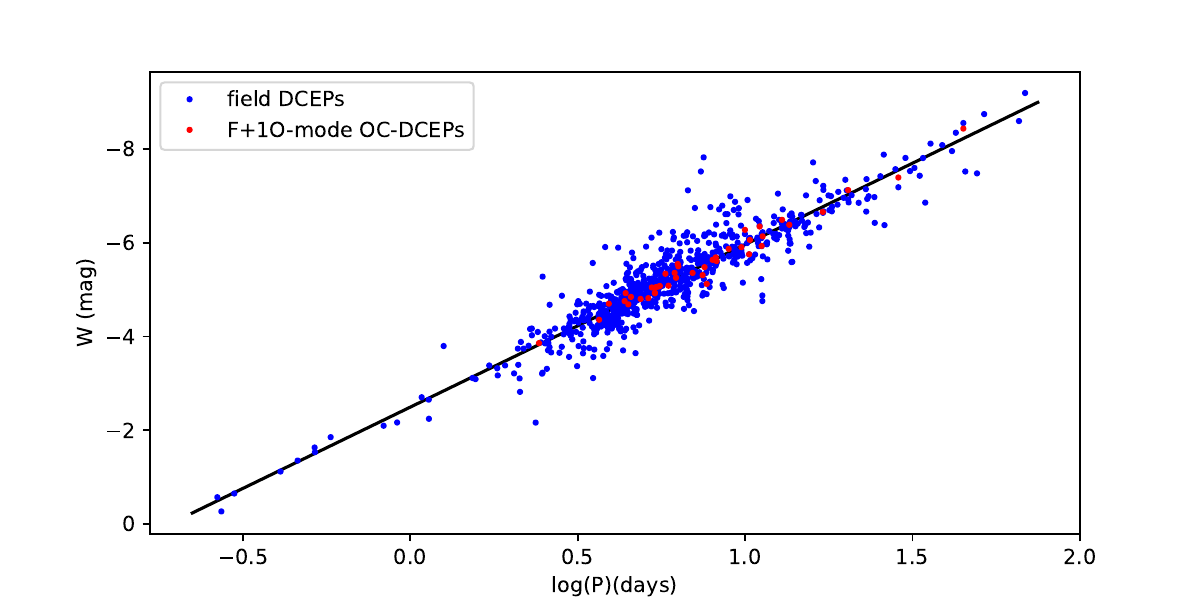}{0.6\textwidth}{}}
\caption{Blue dots are the field DCEPs in \cite{trentin24}. Red dots are our 42 F+1O-mode OC-DCEPs. The black line is the fitting result of our F+1O-mode PWR.
\label{fig:3}}
\end{figure}

\begin{figure}[htb]
\gridline{\fig{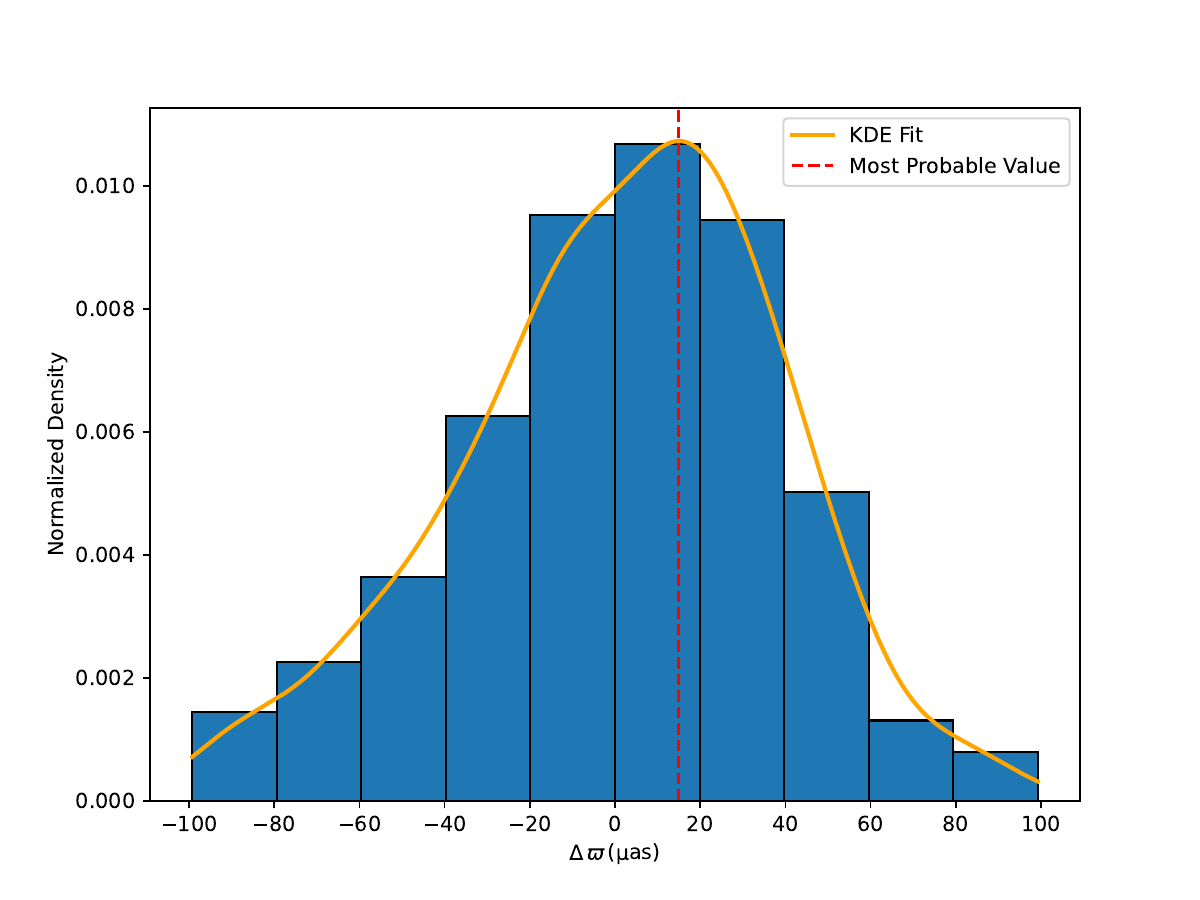}{0.6\textwidth}{}}
\caption{Normalized histogram of parallax offset ($\Delta \varpi$) estimated using our F+1O mode PWZR. The orange curve represents the Gaussian kernel density estimation for this distribution. The red dashed line represents the highest probability density of $\Delta \varpi$.}
\label{fig:4}
\end{figure}

\begin{figure}[htb]
\gridline{\fig{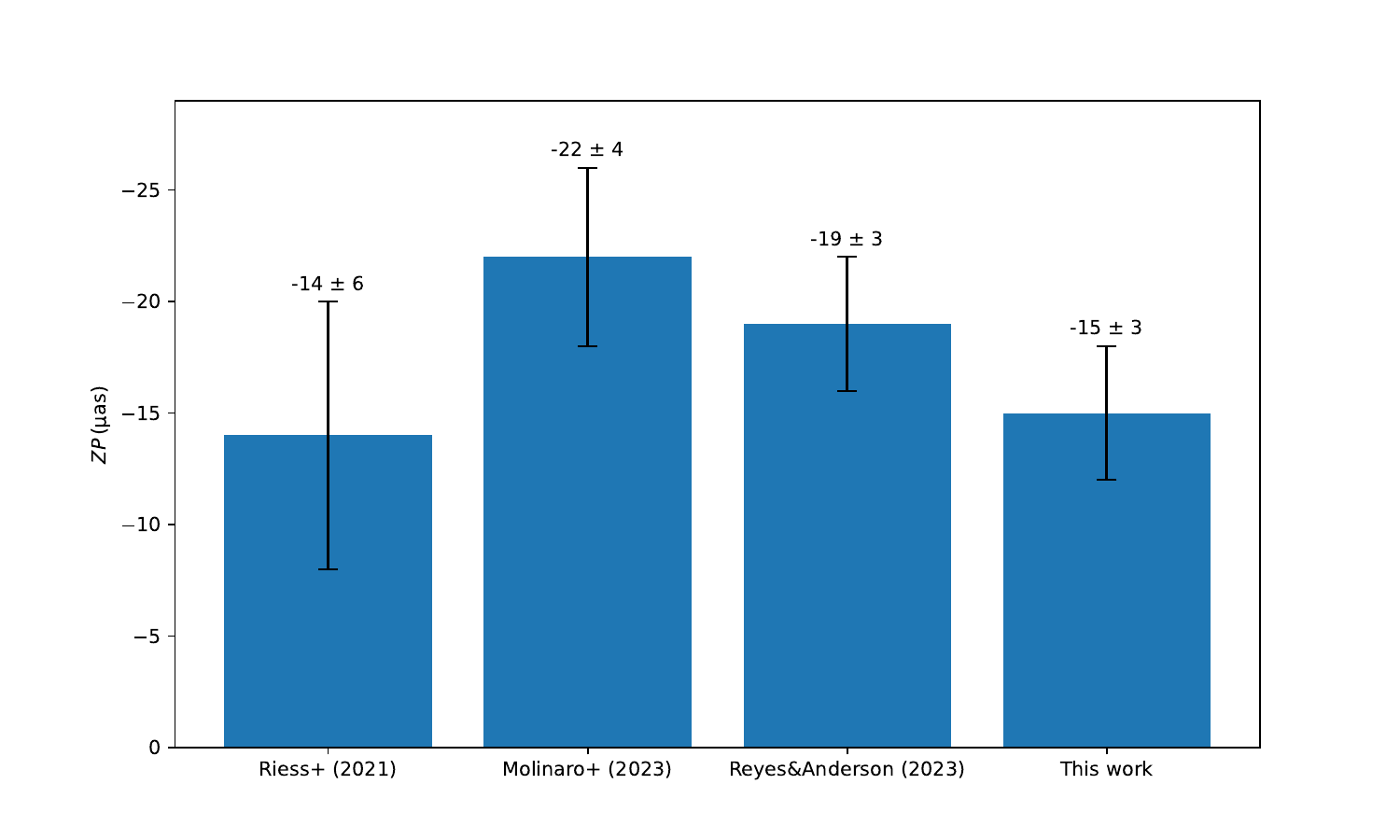}{0.6\textwidth}{}}
\caption{Literature estimates of the $zp$ in field DCEPs.
\label{fig:5}}
\end{figure}

We chose 758 F+1O-mode Galactic field DCEPs from \cite{trentin24} using the following criteria: (1) RUWE \textless 1.4; and (2) $\varpi/\sigma_{\varpi}$ \textgreater 5. The OC-DCEPs we obtained and the field DCEPs are plotted together in Figure \ref{fig:3}. It can be seen that the linear relation between the two is consistent, and the linear relation of OC-DCEPs is tighter than that of field DCEPs. Then, we applied our F+1O-mode PWZR (i.e., Case 3 in Table \ref{tab:1}) on the field DCEPs to derive their photometric parallaxes and parallax offsets, $\Delta \varpi$, between the photometric parallaxes and DR3 parallaxes after L21 corrections (see the distribution of $\Delta \varpi$ in Figure \ref{fig:4}). We convoluted this distribution using a Gaussian kernel density estimate (see the orange curve in Figure \ref{fig:4}) with a bandwidth chosen according to \cite{silver86}, and the $\Delta \varpi$ with the highest probability density (see the red dashed line in Figure \ref{fig:4}) is the estimate of the $zp$ in field DCEPs. To estimate the error of the $zp$ in field DCEPs, we performed 10,000 Monte Carlo simulations, and for each simulation, we randomly varied the coefficients of PWZR ($\alpha$, $\beta$, and $\gamma$) within the error to obtain 10,000 estimates of the $zp$ in field DCEPs, and then calculated their standard deviation as the error. Finally, we obtained an estimate of the $zp$ in field DCEPs as $-15\pm3\,\mu\textrm{as}$, indicating that L21 overcorrects $15\,\mu\textrm{as}$ for field DCEPs. We show Figure \ref{fig:5} to facilitate comparison of the reported $zp$ in field DCEPs. It can be seen that our estimate of the $zp$ in field DCEPs agrees well with that of \cite{riess21}, who estimated the $zp$ in field DCEPs as $-14\pm6\,\mu\textrm{as}$. Our estimate of the $zp$ in field DCEPs is also consistent with other works \citep{molinaro23,reyes23} within 3$\sigma$.

\subsection{Reliability Testing of PWZR on LMC Field DCEPs} \label{sec:LMC}

The PWZR can be used to measure the distances to LMC field DCEPs and thus infer the distance modulus of LMC ($\mu_{\textrm{LMC}}$). One of the most accurate published distance modulus measurements is $\mu_{\textrm{LMC}} = 18.477 \pm 0.004$ (statistical error) $\pm 0.026$ (systematic error) mag, obtained from geometric measurements of eclipsing binaries \citep{pie19}. By comparing our PWZR-based distance modulus with the published $\mu_{\textrm{LMC}}$, it is possible to test the reliability of our PWZR on LMC field DCEPs.

The DCEPs in the LMC were obtained by the following steps. First, we extracted the $\alpha$, $\delta$, and $P$ values of 4525 DCEPs in the LMC from the Optical Gravitational Lensing Survey (OGLE) IV survey \citep{udalski18}. Second, we matched the $\alpha$ and $\delta$ values of each object with gaiadr3.gaia\_source to obtain their source\_id. Finally, we obtained intensity-averaged magnitudes in the three Gaia bands ($G_{\textrm{BP}}$, $G_{\textrm{RP}}$, and $G$) and calculated the corresponding $w_{G}$.

To calculate $W_{G}$, we used our PWZR and assumed all the DCEPs in the LMC have the same metal abundance $[\textrm{Fe/H]}_{\textrm{LMC}} = -0.409 \pm 0.003$ dex \citep{roman22}. The distance modulus of each DCEP was then calculated as $w_{G}-W_{G}$, and we took the median value as our estimate of $\mu_{\textrm{LMC}}$. To estimate the error in $\mu_{\textrm{LMC}}$, we performed 10,000 Monte Carlo simulations, and for each simulation we randomly varied the coefficients ($\alpha$, $\beta$, and $\gamma$) within their errors to obtain 10,000 medians, and then calculated their standard deviation, $\sigma$, as the error of $\mu_{\textrm{LMC}}$.

We list the derived $\mu_{\textrm{LMC}}$ in Table \ref{tab:1}. $\Delta\mu_{\textrm{LMC}}$ represents the difference between $\mu_{\textrm{LMC}}$ measured by PWZR and $\mu_{\textrm{LMC}}$ measured by \cite{pie19}. It is evident from a comparison of Table \ref{tab:1} that the LMC's distance modulus determined by our PWZR is more accurate than found with our PWR, indicating that the latter is indeed affected by the metal abundances of the calibrating DECPs. The $\Delta\mu_{\textrm{LMC}}$ of Case 3 and Case 4 in Table \ref{tab:1} are 0.005 (0.1$\sigma$) and 0.043 (1.1$\sigma$), respectively, which are consistent with \cite{pie19} within 3$\sigma$, confirming the reliability of our PWZR applied to LMC field DCEPs. We consider Case 3 as the optimal PWZR in this work because it best matches the result derived by \cite{pie19}. We also list the $\mu_{\textrm{LMC}}$ derived by using other works' PWZR (i.e., Case 5 to Case 12) in Table \ref{tab:1}. All of their $\mu_{\textrm{LMC}}$ are consistent with \cite{pie19} within 3$\sigma$, with the exception of Case 12, which deviates from \cite{pie19} by 3.4$\sigma$.

\section{Conclusions}\label{sec:conclusions}
We obtained a total of 43 OC-DCEPS, which is the largest sample of OC-DCEPs to date. Benefiting from OC's high-precision parallax, we calibrated the PWZR in the Gaia bands and estimated the $zp$ in OCs simultaneously. We found that the $zp$ in OCs is negligible, demonstrating the adequacy of L21 corrections within the magnitude range of OC member stars. For the metallicity term $\gamma$, we obtained that $\gamma = -0.285\,\pm\,0.064$ for the F+1O-mode OC-DCEPs and $\gamma = -0.162\,\pm\,0.070$ for the F-mode OC-DCEPs, which is consistent with the conclusions of \cite{desomma22}. Applying our F+1O model PWZR on field DCEPs and using a Gaussian kernel density estimate, we found that the $zp = -15\pm3\,\mu\textrm{as}$ in field DCEPs, which is in good agreement with \cite{riess21}. Our best PWZR is $W_G=(-3.356 \,\pm\, 0.033) \,\log(P - 1)+(-5.947 \,\pm\, 0.025)+(-0.285 \,\pm\, 0.064)[\textrm{Fe/H}]$. This PWZR estimates a $\mu_{\textrm{LMC}}$ value of $18.482 \,\pm\, 0.040$ mag, which aligns well with the result derived by \cite{pie19} based on the geometric measurements of eclipsing binaries in the LMC. As more OC-DECPs are identified and more precise astrometric data are published in future releases from Gaia, more precise PWZR will likely be obtained.

\appendix

\section{Examples of OC-DCEPs and rejected associations}\label{A}

\begin{figure*}[htb]
\figurenum{A.1}
\gridline{\fig{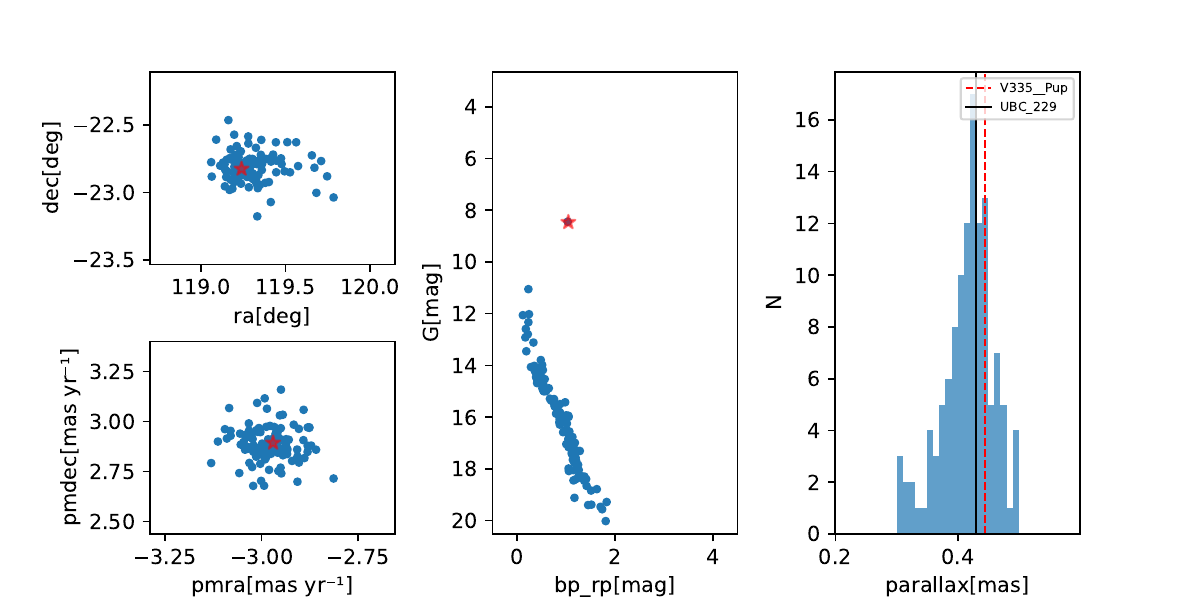}{0.6\textwidth}{(a)}}
\gridline{\fig{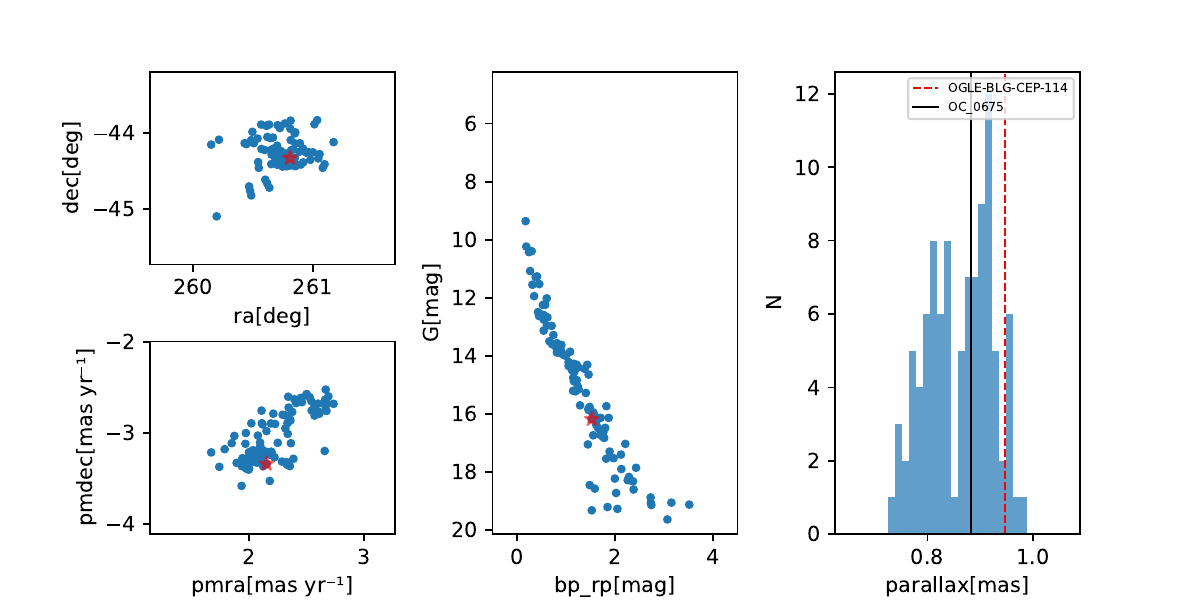}{0.6\textwidth}{(b)}}
\caption{Examples of the OCs (blue dots) harboring one or more DCEPs (red stars). The upper left, lower left, and middle images of each subfigure show their coordinates ($\alpha$ and $\delta$), proper motions ($\mu_{{\alpha}^\ast}$ and $\mu_\delta$) distribution, and the CMD, respectively. The blue histogram on the right shows the parallax distribution of the member stars, where the black solid line represents the mean parallax of the OC and the red dotted line represents the parallax of the DCEP. (a) An example of OC-DCEPs. (b) An example of rejected associations.
\label{fig:6}}
\end{figure*}

Here, we provide an example of our OC-DCEPs as well as an example of rejected associations in figure \ref{fig:6}. Because DCEPs have left the main sequence and entered the instability strip, they should be brighter than the main sequence member stars.

\section{43 Open Cluster Cepheids}\label{B}

Here, we present the parameters for 43 OC-DCEPs obtained by us (listed in Table \ref{tab:2}).

\begin{longrotatetable}
\begin{deluxetable*}{cccccccccccccccccccc}
\setlength{\tabcolsep}{0.6pt}
\tablenum{B.1}
\tablecaption{Parameters of the 43 Open Cluster Cepheids \label{tab:2}}
\tablewidth{0pt}
\tabletypesize{\tiny}
\tablehead{
\multicolumn{9}{c}{Cluster parameter} & \multicolumn{11}{c}{Cepheid parameters}\\
\cline{2-9} \cline{11-20}
\colhead{Cluster} &\colhead{$\alpha$} &\colhead{$\delta$} &\colhead{$\varpi$} &\colhead{$\sigma_{\varpi_{\textrm{OC}}}$} &\colhead{$\mu_{{\alpha}^\ast}$} &\colhead{$\mu_\delta$} &\colhead{$N$} &\colhead{Sep} &\colhead{Cepheid} &\colhead{$\alpha$} &\colhead{$\delta$} &\colhead{$\varpi$} &\colhead{$\mu_{{\alpha}^\ast}$} &\colhead{$\mu_\delta$} &\colhead{$w_G$} &\colhead{Mode} &\colhead{$P_ {\textrm{F}}$} &\colhead{[Fe/H]} &\colhead{Ref} \\
\colhead{} &\colhead{(deg)} &\colhead{(deg)} &\colhead{(mas)} &\colhead{(mas)} &\colhead{($mas\,yr^{-1}$)} &\colhead{($mas\,yr^{-1}$)} &\colhead{} &\colhead{(pc)} &\colhead{} &\colhead{(deg)} &\colhead{(deg)} &\colhead{(mas)} &\colhead{($mas\,yr^{-1}$)} &\colhead{($mas\,yr^{-1}$)} &\colhead{(mag)} &\colhead{} &\colhead{(day)} &\colhead{(dex)} &\colhead{} }
%\decimalcolnumbers
\startdata
         NGC\_7790 & 359.620  & 61.208  & 0.320(0.025) & 0.007  & -3.243(0.059) & -1.726(0.059) & 143 & 21.40  & CG\_\_\_\_Cas & 0.247  & 60.959  & 0.296  & -3.241  & -1.673  & 7.721  & F & 4.365  & 0.06  & G18 \\ 
        NGC\_7790 & 359.620  & 61.208  & 0.320(0.025) & 0.007  & -3.243(0.059) & -1.726(0.059) & 143 & 2.18  & CE\_\_Cas\_B & 359.538  & 61.214  & 0.333  & -3.301  & -1.809  & 7.790  & F & 4.479  & ~ & ~ \\ 
        NGC\_7790 & 359.620  & 61.208  & 0.320(0.025) & 0.007  & -3.243(0.059) & -1.726(0.059) & 143 & 2.14  & CE\_\_Cas\_A & 359.539  & 61.214  & 0.332  & -3.298  & -1.873  & 7.660  & F & 5.141  & ~ & ~ \\ 
        NGC\_7790 & 359.620  & 61.208  & 0.320(0.025) & 0.007  & -3.243(0.059) & -1.726(0.059) & 143 & 1.39  & CF\_\_\_\_Cas & 359.575  & 61.221  & 0.316  & -3.240  & -1.766  & 7.675  & F & 4.875  & -0.01  & G18 \\ 
        SAI\_4 & 5.905  & 62.708  & 0.341(0.033) & 0.008  & -3.098(0.081) & -0.624(0.057) & 71 & 17.80  & V824\_\_Cas & 5.630  & 63.033  & 0.296  & -2.868  & -0.588  & 7.213  & 1O & 7.684  & -0.08  & R21 \\ 
        NGC\_103 & 6.311  & 61.326  & 0.316(0.053) & 0.007  & -2.815(0.110) & -1.066(0.091) & 418 & 7.77  & NO\_\_\_\_Cas & 6.019  & 61.342  & 0.298  & -2.828  & -1.208  & 8.144  & 1O & 3.664  & -0.06  & GDR3 \\ 
        NGC\_129 & 7.590  & 60.206  & 0.559(0.044) & 0.007  & -2.586(0.110) & -1.177(0.108) & 561 & 23.47  & V379\_\_Cas & 6.650  & 60.798  & 0.524  & -2.696  & -1.313  & 5.907  & 1O & 6.162  & 0.12  & G18 \\ 
        NGC\_129 & 7.590  & 60.206  & 0.559(0.044) & 0.007  & -2.586(0.110) & -1.177(0.108) & 561 & 1.51  & DL\_\_\_\_Cas & 7.494  & 60.212  & 0.580  & -2.706  & -1.189  & 5.629  & F & 8.000  & 0.05  & G18 \\ 
        COIN-Gaia\_36 & 36.341  & 59.935  & 0.476(0.042) & 0.007  & -0.985(0.096) & -0.545(0.091) & 183 & 24.41  & GM\_\_\_\_Cas & 36.736  & 60.571  & 0.418  & -0.904  & -0.383  & 6.304  & F & 7.468  & -0.13  & G18 \\ 
        CWNU\_2490 & 58.992  & 55.335  & 0.337(0.043) & 0.009  & -0.442(0.065) & -0.717(0.088) & 30 & 23.47  & MN\_\_\_\_Cam & 59.374  & 54.938  & 0.366  & -0.263  & -0.645  & 6.674  & F & 8.173  & -0.02  & G18 \\ 
        UBC\_1273 & 74.936  & 40.821  & 0.285(0.032) & 0.010  & -0.380(0.067) & -1.502(0.069) & 32 & 1.08  & AN\_\_\_\_Aur & 74.923  & 40.836  & 0.285  & -0.427  & -1.515  & 6.974  & F & 10.289  & -0.13  & G18 \\ 
        OC\_0301 & 88.378  & 25.183  & 0.358(0.061) & 0.008  & -0.400(0.123) & -1.847(0.146) & 100 & 24.06  & J055122+2516.9 & 87.844  & 25.281  & 0.362  & 0.609  & -1.627  & 8.373  & 1O & 2.423  & -0.24  & GDR3 \\ 
        FSR\_0951 & 95.544  & 14.658  & 0.609(0.038) & 0.007  & -0.226(0.103) & -0.013(0.101) & 221 & 0.67  & RS\_\_\_\_Ori & 95.555  & 14.678  & 0.589  & 0.196  & 0.005  & 5.598  & F & 7.567  & 0.11  & G18 \\ 
        vdBergh\_1 & 99.273  & 3.079  & 0.587(0.062) & 0.008  & -0.388(0.113) & -0.711(0.112) & 86 & 0.46  & CV\_\_\_\_Mon & 99.270  & 3.064  & 0.601  & 0.349  & -0.666  & 6.236  & F & 5.379  & 0.09  & G18 \\ 
        UBC\_231 & 115.544  & -26.313  & 0.360(0.033) & 0.007  & -2.247(0.085) & -2.260(0.097) & 98 & 21.25  & WX\_\_\_\_Pup & 115.496  & -25.876  & 0.387  & -2.164  & 2.559  & 6.350  & F & 8.936  & -0.01  & G18 \\ 
        UBC\_1429 & 118.566  & -37.005  & 0.320(0.026) & 0.008  & -2.753(0.084) & -3.770(0.112) & 58 & 5.89  & V724\_\_Pup & 118.438  & -36.970  & 0.318  & -2.678  & 3.857  & 7.400  & F & 5.564  & ~ & ~ \\ 
        UBC\_229 & 119.253  & -22.813  & 0.430(0.041) & 0.008  & -2.979(0.057) & -2.885(0.082) & 121 & 0.70  & V335\_\_Pup & 119.240  & -22.825  & 0.443  & -2.970  & 2.894  & 6.478  & 1O & 6.969  & 0.06  & G18 \\ 
        Ruprecht\_79 & 145.261  & -53.834  & 0.276(0.043) & 0.006  & -4.588(0.076) & -3.043(0.084) & 316 & 1.65  & CS\_\_\_\_Vel & 145.293  & -53.816  & 0.272  & -4.567  & 3.131  & 7.713  & F & 5.905  & 0.09  & G18 \\ 
        HSC\_2354 & 158.759  & -59.631  & 0.235(0.028) & 0.011  & -5.495(0.080) & -2.401(0.062) & 46 & 0.64  & 5254518760118884864 & 158.775  & -59.635  & 0.239  & -5.563  & 2.466  & 8.214  & 1O & 5.379  & -0.18  & GDR3 \\ 
        CWNU\_175 & 188.407  & -63.515  & 0.730(0.032) & 0.006  & -3.901(0.106) & -1.186(0.086) & 55 & 0.87  & VW\_\_\_\_Cru & 188.328  & -63.506  & 0.738  & -3.903  & -1.134  & 5.638  & F & 5.265  & 0.16  & G18 \\ 
        UBC\_290 & 191.742  & -59.376  & 0.644(0.039) & 0.007  & -5.952(0.106) & -0.213(0.103) & 355 & 7.12  & X\_\_\_\_\_Cru & 191.593  & -59.125  & 0.654  & -5.926  & -0.173  & 5.704  & F & 6.220  & 0.12  & G18 \\ 
        NGC\_5662 & 218.927  & -56.575  & 1.332(0.050) & 0.007  & -6.495(0.180) & -7.204(0.186) & 439 & 7.00  & V\_\_\_\_\_Cen & 218.138  & -56.888  & 1.409  & -6.697  & -7.068  & 4.317  & F & 5.494  & 0.12  & G18 \\ 
        CWNU\_19 & 228.514  & -54.536  & 0.536(0.033) & 0.007  & -0.780(0.120) & -1.758(0.101) & 84 & 9.21  & IQ\_\_\_\_Nor & 228.206  & -54.755  & 0.535  & -0.897  & -1.821  & 5.747  & F & 8.220  & 0.22  & G18 \\ 
        Theia\_3005 & 242.885  & -54.321  & 0.512(0.018) & 0.008  & -1.882(0.124) & -3.862(0.112) & 97 & 1.50  & QZ\_\_\_\_Nor & 242.835  & -54.354  & 0.484  & -1.896  & -3.848  & 6.403  & 1O & 5.407  & 0.21  & G18 \\ 
        NGC\_6067 & 243.295  & -54.232  & 0.511(0.038) & 0.007  & -1.961(0.118) & -2.578(0.119) & 1149 & 0.56  & V340\_\_Nor & 243.322  & -54.235  & 0.491  & -2.066  & -2.634  & 5.323  & F & 11.289  & 0.07  & G18 \\ 
        NGC\_6087 & 244.683  & -57.914  & 1.066(0.054) & 0.007  & -1.601(0.202) & -2.427(0.163) & 360 & 0.37  & S\_\_\_\_\_Nor & 244.716  & -57.900  & 1.099  & -1.608  & -2.136  & 3.956  & F & 9.754  & 0.10  & G18 \\ 
        UBC\_1558 & 252.706  & -45.414  & 0.429(0.028) & 0.008  & -1.319(0.097) & -2.503(0.073) & 64 & 5.87  & KQ\_\_\_\_Sco & 252.911  & -45.427  & 0.472  & -1.366  & -2.497  & 4.450  & F & 28.703  & 0.52  & G18 \\ 
        HSC\_2961 & 267.432  & -32.977  & 0.696(0.032) & 0.008  & -1.879(0.187) & -1.776(0.185) & 51 & 19.22  & RY\_\_\_\_Sco & 267.718  & -33.706  & 0.764  & 1.485  & -1.388  & 3.665  & F & 20.322  & 0.01  & G18 \\ 
        CWNU\_1841 & 272.828  & -20.884  & 0.402(0.016) & 0.013  & -0.033(0.174) & -1.452(0.125) & 41 & 11.02  & VY\_\_\_\_Sgr & 273.019  & -20.704  & 0.412  & 0.307  & -1.548  & 5.598  & F & 13.558  & 0.33  & G18 \\ 
        IC\_4725 & 277.942  & -19.131  & 1.551(0.045) & 0.007  & -1.692(0.188) & -6.165(0.218) & 725 & 0.32  & U\_\_\_\_\_Sgr & 277.972  & -19.125  & 1.605  & -1.795  & -6.127  &   & F & 6.745  & 0.14  & G18 \\ 
        NGC\_6649 & 278.359  & -10.402  & 0.510(0.063) & 0.007  & -0.037(0.131) & -0.115(0.133) & 728 & 1.55  & V367\_\_Sct & 278.397  & -10.427  & 0.473  & 0.082  & -0.273  & 5.916  & F1O & 6.293  & 0.05  & G18 \\ 
        NGC\_6664 & 279.118  & -8.206  & 0.502(0.054) & 0.007  & -0.099(0.158) & -2.593(0.151) & 482 & 1.78  & EV\_\_\_\_Sct & 279.165  & -8.185  & 0.526  & -0.209  & -2.546  & 6.575  & 1O & 4.398  & 0.09  & G18 \\ 
        CWNU\_337 & 279.158  & -8.909  & 0.570(0.034) & 0.009  & -0.731(0.071) & -2.846(0.086) & 32 & 19.72  & Y\_\_\_\_\_Sct & 279.514  & -8.369  & 0.558  & -0.737  & -2.878  & 5.162  & F & 10.341  & 0.20  & G18 \\ 
        Trumpler\_35 & 280.747  & -4.228  & 0.374(0.051) & 0.008  & -0.983(0.090) & -2.243(0.107) & 257 & 10.43  & TY\_\_\_\_Sct & 280.533  & -4.293  & 0.371  & -1.106  & -2.466  & 5.791  & F & 11.054  & 0.34  & G18 \\ 
        Trumpler\_35 & 280.747  & -4.228  & 0.374(0.051) & 0.008  & -0.983(0.090) & -2.243(0.107) & 257 & 7.38  & CN\_\_\_\_Sct & 280.627  & -4.331  & 0.390  & -1.042  & -2.255  & 5.862  & F & 9.994  & 0.30  & G18 \\ 
        UBC\_106 & 280.492  & -5.417  & 0.440(0.041) & 0.007  & -1.053(0.099) & -1.361(0.115) & 664 & 5.61  & CM\_\_\_\_Sct & 280.612  & -5.341  & 0.444  & -1.064  & -1.414  & 7.084  & F & 3.917  & 0.12  & G18 \\ 
        NGC\_6683 & 280.566  & -6.225  & 0.328(0.032) & 0.008  & -0.343(0.067) & -2.344(0.063) & 83 & 23.32  & Z\_\_\_\_\_Sct & 280.739  & -5.821  & 0.357  & -0.379  & -2.205  & 5.931  & F & 12.902  & 0.12  & G18 \\ 
        UBC\_130 & 298.061  & 27.449  & 0.424(0.027) & 0.007  & -2.107(0.061) & -5.876(0.108) & 142 & 6.67  & SV\_\_\_\_Vul & 297.879  & 27.460  & 0.402  & -2.158  & -5.962  & 3.429  & F & 44.894  & 0.11  & G18 \\ 
        UBC\_129 & 299.035  & 26.445  & 0.887(0.056) & 0.007  & -0.984(0.106) & -4.369(0.125) & 348 & 6.29  & X\_\_\_\_\_Vul & 299.369  & 26.556  & 0.864  & -1.352  & -4.247  & 4.766  & F & 6.320  & 0.13  & G18 \\ 
        UBC\_135 & 299.817  & 33.724  & 0.267(0.043) & 0.007  & -3.512(0.069) & -6.429(0.099) & 163 & 4.24  & GI\_\_\_\_Cyg & 299.890  & 33.746  & 0.273  & -3.452  & -6.577  & 7.527  & F & 5.783  & 0.24  & G18 \\ 
        Berkeley\_84 & 301.200  & 33.986  & 0.393(0.043) & 0.008  & -2.012(0.070) & -5.555(0.096) & 103 & 6.51  & CD\_\_\_\_Cyg & 301.111  & 34.112  & 0.394  & -1.970  & -5.583  & 5.370  & F & 17.079  & 0.12  & G18 \\ 
        vdBergh\_130 & 304.517  & 39.367  & 0.596(0.035) & 0.006  & -3.548(0.203) & -5.127(0.165) & 162 & 20.97  & V438\_\_Cyg & 304.726  & 40.064  & 0.530  & -3.324  & -4.559  & 5.193  & F & 11.210  & 0.30  & G18 \\ 
        Kronberger\_84 & 323.888  & 53.514  & 0.210(0.025) & 0.008  & -2.920(0.087) & -3.032(0.065) & 79 & 0.12  & J213533.70+533049.3 & 323.890  & 53.514  & 0.214  & -2.878  & -3.113  & 8.541  & 1O & 4.561  & -0.12  & GDR3 \\ 
\enddata
\tablecomments{Values in parentheses are standard deviations. $N$ is the number of member stars in the OC. $\sigma_{\varpi_{\textrm{OC}}}$ is the total uncertainty, including the contribution from angular covariance. “Sep” is the distance between the DCEP and OC. $w_G$ is the apparent Wesenheit magnitude in the Gaia bands. “Ref” is the reference for the matallicity of DCEP.}
\tablerefs{G18: \cite{greo18}; R21: \cite{ripepi21}; GDR3: \cite{GDR3}.}
\end{deluxetable*}
\end{longrotatetable}

\section{the marginalised posterior of distributions}\label{C}

\begin{figure*}[htb]
\figurenum{C.1}
\gridline{\fig{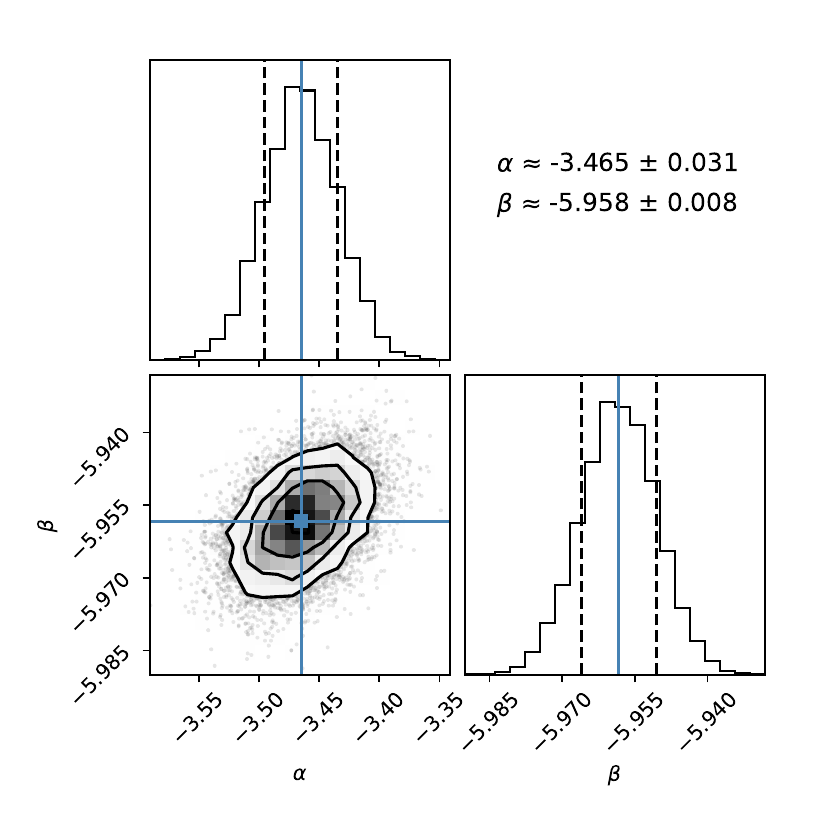}{0.5\textwidth}{} \hfill \fig{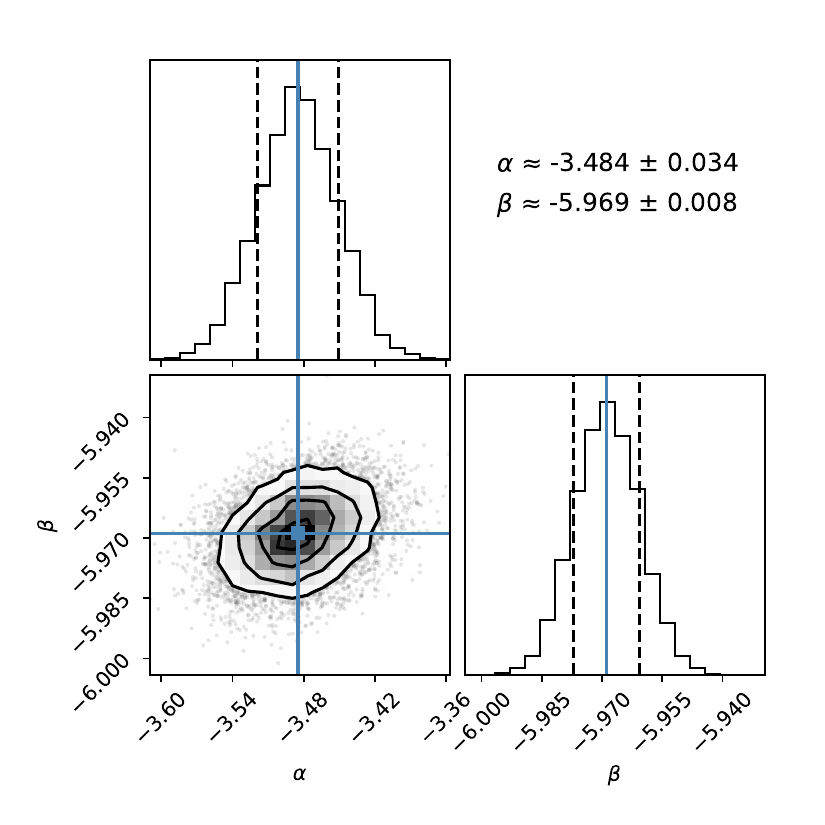}{0.5\textwidth}{}}
\gridline{\fig{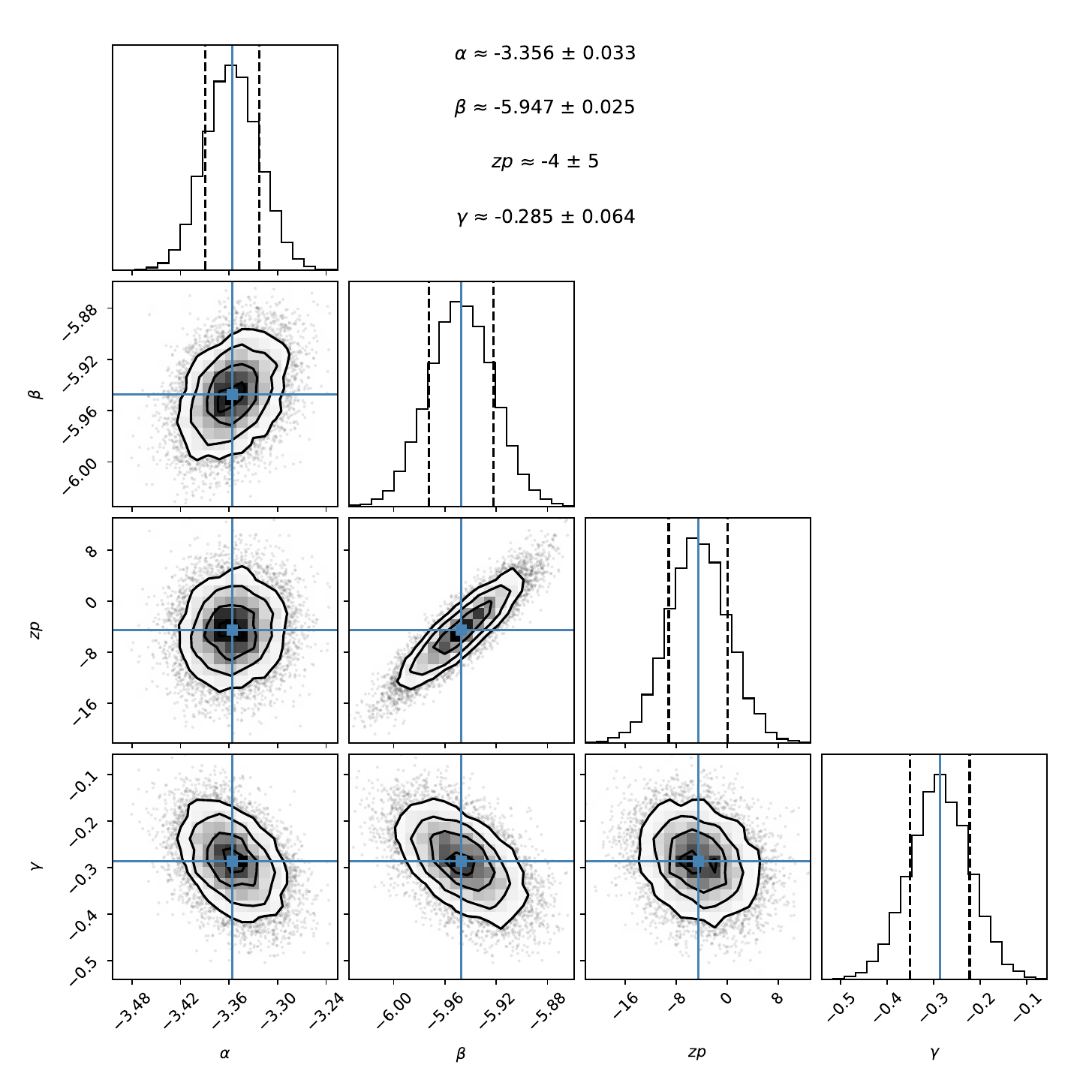}{0.5\textwidth}{} \hfill \fig{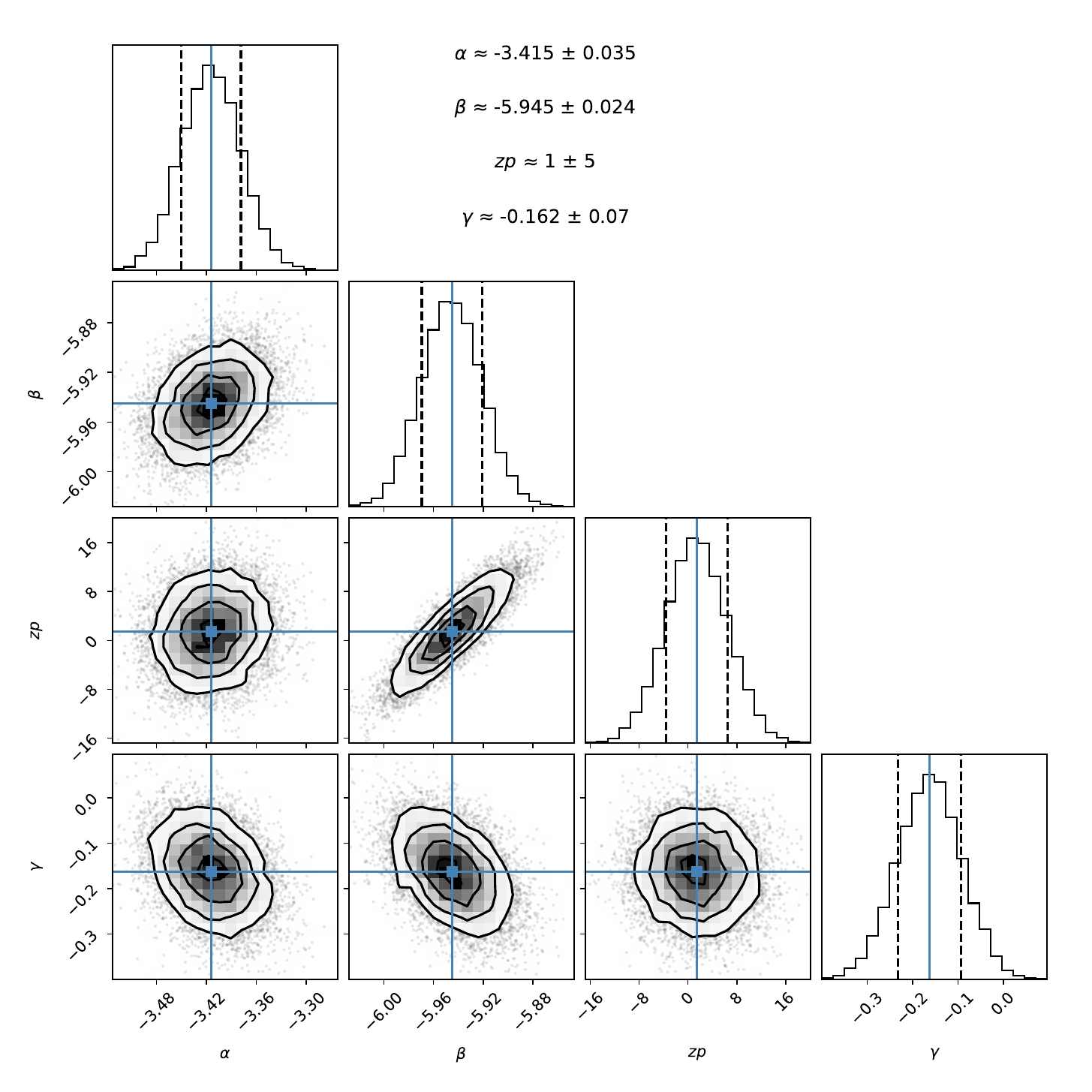}{0.5\textwidth}{}}
\caption{The upper and lower corner plots represent the marginalised posterior distributions of the free parameters in PWR and PWZR, respectively. The vertical blue lines represent the median values, and the black dashed lines represent the 16th and 84th percentiles.
\label{fig:7}}
\end{figure*}

%\bibliography{apj/sample631}{}
%\bibliographystyle{aasjournal}


\end{CJK}
\end{document}